\documentclass[superscriptaddress,twocolumn,prl]{revtex4}

\usepackage[T1]{fontenc}
\usepackage[utf8]{inputenc}

\usepackage{amsmath}
\usepackage{amsfonts}
\usepackage{amssymb}
\usepackage{mathtools}
\usepackage{graphicx}
\graphicspath{{images/}}
\usepackage{color}
\usepackage[pdfstartview=FitH,
            breaklinks=true,
            bookmarksopen=false,
            bookmarksnumbered=true,
            colorlinks=true,
            linkcolor=black,
            citecolor=black,
            urlcolor=black,
            pdftitle={Unbiased estimation of sampling variance for Simpson's diversity index},
            pdfauthor={Andreas Tiffeau-Mayer},
            pdfsubject={Uncertainty quantification in biodiversity measurements}
            ]{hyperref}

\newcommand{\<}{\langle}
\renewcommand{\>}{\rangle}
\newcommand{\Var}{\mathrm{Var}}

\def\(({\left(}
\def\)){\right)}                       
\def\[[{\left[}
\def\]]{\right]}

\begin{document}

\title{Unbiased estimation of sampling variance for Simpson's diversity index}
\author{Andreas Tiffeau-Mayer}
\affiliation{Division of Infection and Immunity \& Institute for the Physics of Living Systems, University College London}
\date{\today}

\begin{abstract}
Quantification of measurement uncertainty is crucial for robust scientific inference, yet accurate estimates of this uncertainty remain elusive for ecological measures of diversity. Here, we address this longstanding challenge by deriving a closed-form unbiased estimator for the sampling variance of Simpson's diversity index. In numerical tests the estimator consistently outperforms existing approaches, particularly for applications in which species richness exceeds sample size. We apply the estimator to quantify biodiversity loss in marine ecosystems and to demonstrate ligand-dependent contributions of T cell receptor chains to specificity, illustrating its versatility across fields. The novel estimator provides researchers with a reliable method for comparing diversity between samples, essential for quantifying biodiversity trends and making informed conservation decisions.
\end{abstract}

\maketitle

Living systems are characterized by immense diversity across multiple scales from molecules to ecosystems \cite{Allesina2012StabilityCriteria,Posfai2017MetabolicTradeOffs,Pearce2020StabilizationExtensive,Marcou2018HighThroughputImmune,Sethna2019OLGAFast}. Quantitatively understanding how this diversity is produced and supports biological function has been a central question in the physics of living systems: On the ecosystem scale, statistical physics approaches have for instance identified conditions under which diverse interacting species can be stably maintained \cite{Allesina2012StabilityCriteria,Posfai2017MetabolicTradeOffs,Pearce2020StabilizationExtensive}, while on the molecular and cellular scale, probabilistic modelling has shed light on how antibody and T cell receptor diversity in the adaptive immune system are generated \cite{Marcou2018HighThroughputImmune,Sethna2019OLGAFast,Mayer2015HowWellAdapted,Mayer2019HowWelladapting}. 

To compare predictions from ecological theory to experimentally measured diversities requires quantification of measurement uncertainty; similarly comparing changes in biodiversity in response to habitat loss or climate change requires determining whether diversity differs more than expected by sampling chance alone \cite{Haegeman2013RobustEstimation,Roswell2021ConceptualGuide,Willis2019RarefactionAlpha,Albano2021NativeBiodiversity}. A number of estimators have been proposed to quantify sampling variance in diversity estimation \cite{Simpson1949MeasurementDiversity,Nemenman2001EntropyInference,Grundmann2001DeterminingConfidence,Chao2014RarefactionExtrapolation,Kaplinsky2016RobustEstimates}. However, none of the current estimators are unbiased outside of asymptotically large samples and numerical tests suggests that their finite sample bias can be severe. This is a practically important limitation, as overestimation of sampling uncertainty diminishes the ability to detect true trends in biodiversity. Reversely, underestimating sampling variance can inflate the apparent significance of observed changes in diversity, potentially leading to spurious conclusions.

Here, we address this gap by deriving an unbiased estimator of the sampling variance of Simpson's index. We show the superior performance of the estimator compared to previous approaches on simulated and real data and apply it to establish ligand-dependent differences in T cell receptor diversity. Simpson's index is used widely from ecology \cite{Simpson1949MeasurementDiversity,Jost2006EntropyDiversity} and microbiology \cite{Hunter1988NumericalIndex,Grundmann2001DeterminingConfidence} to economics \cite{Gibbs1962UrbanizationTechnology} as a measure of species diversity.
 The index is defined as the probability of species coincidence of a random pair of individuals,
\begin{equation} \label{eqdefpc}
    p_C = \sum_{i=1}^S p_i^2,
\end{equation}
where $p_i$ is the frequency of species $i$ in the population  $i=1, ..., S$ and $S$ the number of distinct species. This index can be converted into an effective number of species $D = 1/p_C$ \cite{Jost2006EntropyDiversity}. Importantly, the unbiased estimator introduced here provides for the first time interval estimates for diversity that do not depend systematically on sample size. It thus addresses a recently highlighted gap in methods \cite{Willis2019RarefactionAlpha} to compare diversity estimates between samples of different sizes without rarefaction.

\begin{figure}[b]
\begin{center}
    \includegraphics{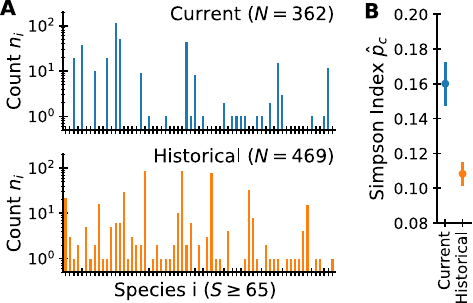}
    \caption{{\bf Quantifying collapse of marine biodiversity in the Eastern Mediterranean sea.} Illustration of the problem setting of uncertainty quantification for finite sample diversity estimates. (A) Distributions of sampled mollusc species currently alive (top) and found in surficial dead assemblages (bottom). (B) Simpson Diversity index with error bars calculated using the proposed method (Eq.~\ref{eqhatvarhatpc}). Error bars show $\hat{p_C} \pm \sqrt{\hat{\Var}(\hat{p_C})}$ and demonstrate statistical significance of the difference in diversity. Data: Non-invasive mollusc species found on shallow subtidal ground at 12m depth of the coast of Ashqelon \cite{Albano2021NativeBiodiversity}.}
    \label{figillustration}
\end{center}      
\end{figure}

To illustrate the problem setting we reanalyze data from \citet{Albano2021NativeBiodiversity} (Fig.~1), who considered the following question: Has climate change led to a biodiversity loss in the Mediterranean sea? Fig.~\ref{figillustration}A shows the raw data, the number $n_i$ of counted molluscs belonging to species $i$ in a patch on the sea floor (top) and corresponding counts for empty mollusc shells found at the same site in death assemblages (bottom). A diversity index turns these species counts into scalar measures of current and past biodiversity (points in Fig.~\ref{figillustration}B). However, less than 500 molluscs were counted across 65 distinct species, so some sampling variability in the diversity estimates is expected. The method we introduce allows robust quantification of this sampling variance (error bars in Fig.~\ref{figillustration}B). Here addition of confidence intervals rules out sampling variability as the sole source of the differences supporting the conclusion that biodiversity has decreased significantly.

\section{The unbiased estimator}
In 1949 Simpson published a short letter in Nature that proposed the index that now bears his name \cite{Simpson1949MeasurementDiversity}. In the same publication Simpson also showed that 
\begin{equation} \label{eqhatpc}
    \hat{p_C} = \frac{\sum_{i=1}^S n_i (n_i-1) }{N(N-1)}
\end{equation}
provides an unbiased estimate of underlying population diversity from a finite sample of size $N$. Here $n_i$ with $\sum_i^S n_i = N$ is the number of counts of the $i$-th species in the sample, which follows a multinomial distribution 
\begin{equation} \label{eqmultinomial}
    P(n_1, \dots, n_S) = \frac{N!}{\prod_{i=1}^S n_i!} \prod_{i=1}^S p_i^{n_i}
\end{equation}
under the commonly used assumption that each individual from the population is sampled with equal probability \cite{Simpson1949MeasurementDiversity,Grundmann2001DeterminingConfidence,Chao2014RarefactionExtrapolation}.

Here we propose that the variance $\Var(\hat{p_C})$ of the point estimate (Eq.~\ref{eqhatpc}) can be calculated without bias using the following closed-form estimator:
\begin{equation} \label{eqhatvarhatpc}
    \hat{\Var}(\hat{p_C})= \frac{a}{1-b} \hat p_T - \frac{b}{1-b} \hat{p_C}^2 + \frac{c}{1-b} \hat{p_C},
\end{equation}
where
\begin{equation} \label{eqhatpT}
    \hat{p_T} = \frac{\sum_{i=1}^S n_i (n_i-1) (n_i-2) }{N(N-1)(N-2)}
\end{equation}
and
\begin{equation} \label{eqabc}
    a = \frac{4(N-2)}{N(N-1)}, \;
    b = \frac{2(2N-3)}{N(N-1)}, \;
    c = \frac{2}{N(N-1)}.
\end{equation}

\section{Background and derivation}

For the following derivation, it is instructive to recall the insight leading to the unbiased point estimator given in Eq.~\ref{eqvarhatpc}. To estimate $p_C$ from a sample, it is tempting to simply replace the population frequencies $p_i$ in Eq.~\ref{eqdefpc} with the sampled frequencies $f_i = n_i/N$. However, one is well advised to resist this temptation as such plugin estimators are known to be severely biased in small samples \cite{Simpson1949MeasurementDiversity,Nemenman2001EntropyInference,Grassberger2003EntropyEstimates,Chao2014RarefactionExtrapolation}. Instead Eq.~\ref{eqhatpc} should be used, which is the probability of coincidence when drawing pairs of items from the sample without replacement. This estimator of $p_C$ is unbiased, i.e. $\langle \hat{p_C} \rangle = p_C$, where $\langle . \rangle$ is an average over repeated samples of a fixed size. Evaluating the expectation requires calculating the factorial moment $\langle n_i (n_i-1) \rangle$, which can be calculated most conveniently from the probability generating function $G(z_1, \dots, z_S) = \langle \prod_i z_i^{n_i} \rangle$. From the definition of $G$ it follows that $\frac{\partial}{\partial z_i^2} G(1, \dots, 1) =  \langle n_i (n_i-1) \rangle$. For the multinomial distribution $G(z_1, \dots, z_S) = (\sum_i p_i z_i)^N$, and thus
\begin{equation} \label{eq_factorial_2nd}
    \langle n_i (n_i-1) \rangle = N(N-1) p_i^2,
\end{equation}
which can be plugged into Eq.~\ref{eqhatpc} to complete the proof that $\hat{p_C}$ is unbiased.

\begin{figure*}
\begin{center}
    \includegraphics{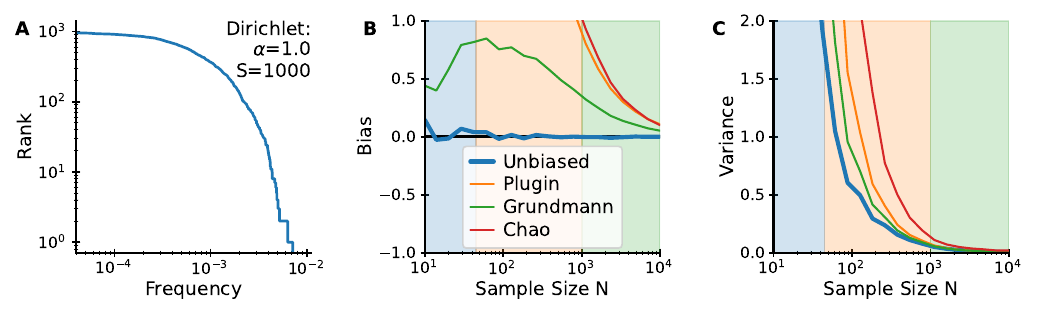}
    \caption{{\bf Benchmarking of variance estimators on simulated data.} (A) Frequency-rank plot of species abundances. Probabilities of $S=1000$ species were drawn from the steady-state Dirichlet distribution of a neutral model with immigration and stochastic drift (parameter $\alpha=1$). (B) Bias and (C) variance as a function of sample size $N$. Expectation values were calculated over 1000 repeated draws at each sample size from the population distribution. Bias and variance are expressed as fractions, i.e. divided by the true value or its square, respectively. The shaded areas differentiate sampling regimes: blue $N < \sqrt{2S}$, orange $\sqrt{2S} < N < S$, and green $N > S$. The number of bootstrap samples for Chao's method was set to 200 as recommended \cite{Chao2014RarefactionExtrapolation}.}
    \label{fignumerics}
\end{center}      
\end{figure*}

We now turn to reviewing the state of the art for variance estimation for Simpson's index. We first recall the formula for the sampling variance of the point estimator was again already given by Simpson \cite{Simpson1949MeasurementDiversity} (derived in the appendix for completeness),
\begin{equation} \label{eqvarhatpc}
\Var(\hat{p_C}) = a p_T - b p_C^2 + c p_C.
\end{equation}
This estimator is a linear combination of the triplet coincidence probability,
\begin{equation}
    p_T = \sum_{i=1}^S p_i^3,
\end{equation}
the square of the coincidence probability $p_C^2$ and $p_C$ with sample size dependent parameters a, b and c given in Eq.~\ref{eqabc}.
Based on this formula, Grundmann et al. \cite{Grundmann2001DeterminingConfidence} proposed estimating variance by plugging the empirical frequencies $f_i = n_i/N$ into an asymptotic expansion of Eq.~\ref{eqvarhatpc} for $N \to \infty$,
\begin{equation}
    \hat{\Var}_{\mathrm{Grundmann}}(\hat{p_C}) = \frac{4}{N} \left(\sum_{i=1}^S  f_i^3 - \left(\sum_{i=1}^S f_i^2\right)^2\right).
\end{equation}
However, we will find that this plugin estimator is substantially biased for small $N$, similarly to plugin estimators of diversity indices themselves. We will also find substantial biases for the non-asymptotic plugin estimator
\begin{equation} \label{eqvarhatpc_plugin}
    \hat{\Var}_{\mathrm{plugin}} = a \sum_{i=1}^S f_i^3 - b \left(\sum_{i=1}^S f_i^2\right)^2 + c \sum_{i=1}^S f_i^2.
\end{equation}
In addition to these plugin estimators, another popular approach is due to \citet{Chao2014RarefactionExtrapolation}. This method is widely used in the field due to its implementation in the R package iNEXT \cite{Hsieh2016INEXTPackage}. Chao's method estimates variances by bootstrapping from a population constructed from the sample by coverage-reweighting observed species frequencies and by augmenting the sample with an estimated number of rare unseen species. Surprisingly, despite its widespread use, we will find that this estimator has the largest bias and variance of tested methods. 

Can we generalize the coincidence counting approach underlying the unbiased point estimate to the problem of interval estimation? 
To derive a better estimator for the variance of Simpson's index we exploit the linearity of Eq.~\ref{eqvarhatpc} and decompose the problem into the unbiased estimation of each of the three terms.
For the first term, the analogy to $\hat{p_C}$ suggests to estimate the triplet probability using Eq.~\ref{eqhatpT}. To show that this estimator is indeed unbiased requires calculating the third factorial moment $\langle n_i (n_i-1) (n_i-2) \rangle$. This factorial moment can again be calculated by taking derivatives of the probability generating function $G$ of the multinomial distribution,
\begin{align}
    \langle n_i (n_i-1) (n_i-2) \rangle &= \frac{\partial^3 G(1, \dots, 1)}{\partial z_i^3} \\
    &=  N(N-1)(N-2) p_i^3,
\end{align}
which can be plugged into the definition of $\hat{p_T}$ to demonstrate its absence of bias.
For the second term, we re-express the squared coincidence probability as
\begin{equation}
    p_C^2 = \langle \hat{p_C}^2 \rangle - \Var(\hat{p_C}).
\end{equation}
where we have used the variance decomposition formula, $\Var(\hat{p_C}) = \langle \hat{p_C}^2 \rangle - \langle \hat{p_C} \rangle^2$, and the unbiasedness of Simpson's estimator, $\langle \hat{p_C} \rangle = p_C$.
Plugging this expression into Eq.~\ref{eqvarhatpc} and solving for $\Var(\hat{p_C})$ yields
\begin{equation}
\Var(\hat{p_C}) = \frac{a}{1-b} p_T - \frac{b}{1-b} \langle \hat{p_C}^2\rangle + \frac{c}{1-b} p_C.
\end{equation}
Finally, the third term can be estimated using Eq.~\ref{eqhatpc}. Combining these results proves our central finding, the unbiasedness of the estimator proposed in Eq.~\ref{eqhatvarhatpc}.

\section{Benchmarking on simulated data}

To compare the empirical performance of the different estimators we turned to numerical experiments, applying estimators to samples from a range of ecologically relevant species abundance distributions. We simulated drawing samples of different sizes $N$ from the population ranging from $N=10$ to $N=10000$.
Repeated sampling at a given sample size allows evaluation of how empirical estimates deviate from the ground truth value, $\Var(\hat p_C)$, computed using Eq.~\ref{eqvarhatpc} from the species abundances.
Given an estimator $\hat x$ of a parameter with true value $x$ a natural measure of its quality is the mean squared error,
\begin{equation}
    \text{MSE}(\hat x) = \< (\hat x - x)^2\>,
\end{equation}
which can be decomposed into a bias and variance term,
\begin{equation}
    \text{MSE}(\hat x) = \text{Bias}(\hat x)^2 + \Var(\hat x),
\end{equation}
where $\text{Bias}(\hat x) = \< \hat x - x \>$ and $\Var(\hat x) = \< (\hat x - \< \hat x \>)^2 \>$. To make values more readily interpretable we normalize Bias and Variance by the true value $x$ to the fractional values, $\text{Bias}/x$ and $\Var/x^2$. We display bias (Fig.~\ref{fignumerics}B) and variance (Fig.~\ref{fignumerics}C) separately, to investigate any potential trade-offs between bias and variance \cite{Mehta2019HighbiasLowvariance}. 

In our first experiments, we drew relative species abundances from Dirichlet distributions, $\rho(p_i) \propto p_i^{\alpha-1}$. These distributions arise in ecology as the steady-state of neutral birth, death and immigration dynamics in a Wright-Fisher diffusion limit \cite{Etheridge2012MathematicalModels,Mayer2019HowWelladapting}. We sampled a species abundance distribution of support $S=1000$ by sampling $p_i$ uniformly form the probability simplex, this is from a Dirichlet distribution with $\alpha=1$ (Fig.~\ref{fignumerics}A). 
The numerical results demonstrate that the proposed estimator is not only unbiased (Fig.~\ref{fignumerics}B), but also has lower variance than all other estimators (Fig.~\ref{fignumerics}C). 
We also generated a more uniform and more peaked distribution of species abundances, corresponding respectively to high or low relative immigration rates, by using a Dirichlet parameter $\alpha=4$ (Fig.~\ref{fignumericsSI}A) and $\alpha=0.25$ (Fig.~\ref{fignumericsSI}D), respectively. We find that the unbiased estimator consistently performs best regardless of the choice of $\alpha$ (Fig.~\ref{fignumericsSI}B-C,E-F).

Different sample size regimes, which we analzye in detail in the next section, predict the performance of the different estimators. For sample sizes $N > S$ (green shading) all estimators have relatively low variance, but Chao's estimator and the full plugin estimator are substantially biased unless $N \gg S$. All estimators are highly variable for very small sample sizes $N < \sqrt{2S}$ (blue shading), which corresponds to the sample size at which the expected number of coincidences for a uniform distribution of $S$ species is one. 

To test the generality of these findings beyond neutral models, we repeated the numerical experiment with alternative species abundance distributions. In theoretical ecology, chaotic competition in generalized Lotka-Volterra models \cite{Pearce2020StabilizationExtensive} and stochastic environmental fluctuations \cite{Gaimann2020EarlyLife} have been shown to lead to the emergence of heavy-tailed species abundance distributions. Empirically, long-standing evidence shows that species abundance distributions in many complex ecosystems are well fit by lognormal \cite{Preston1948CommonnessRarity} or power-law distributions \cite{Yule1924MathematicalTheory}. We thus tested our estimator on samples from species abundance distributions following a lognormal (Fig.~\ref{fignumericslognormal}) and power-law form (Fig.~\ref{fignumericszipf}). The results again show that other methods have substantial bias in small samples that is removed by the proposed estimator.

\section{Insights into sample size scaling}

To gain intuition into how estimator performance depends on $N$ we consider limits of the variance formula (Eq.~\ref{eqvarhatpc}).  For large $N$, the variance is asymptotically equal to
\begin{equation}
    \Var(\hat{p_c}) = \frac{4}{N} \left( \sum_i p_i^3 - \left(\sum_i p_i^2\right)^2 \right).
\end{equation}
This shows that in large samples the variance of the estimator scales with the familiar $1/N$ scaling of an arithmetic average. Note further that the expression in brackets can be interpreted as the variance of species probabilities, $\Var(p) = \< p^2 \> - \<p\>^2 = p_T - p_C^2$. When this variance is estimated by plugging in empirical frequencies, there is an additional sampling variance contribution, which explains why plugin estimators are positively biased.

Conversely, when the number of species $S$ is increased at fixed $N$ the third term in Eq.~\ref{eqvarhatpc} asymptotically dominates as $p_T \sim 1/S^2$ and $p_C^2 \sim 1/S^2$ while $p_C \sim 1/S$, thus
\begin{equation} \label{eqvarhatpc_poisson}
    \Var(\hat{p_c}) = \frac{2}{N(N-1)} p_C.
\end{equation}
Interestingly, the variance scales as $1/N^2$ in this limit, which explains the sharp rise in estimator variances as $N \sim \sqrt{2S}$. As coincidences are rare they occur roughly independently across the $N(N-1)/2$ possible pairs and the distribution of the total number of coincidences $n_C$ is approximately Poissonian \cite{Aldous1989ProbabilityApproximations} with mean $\langle n_C \rangle = N(N-1)p_C/2$. The variance of a Poisson distribution equals its mean, and as $\hat{p_C} = \frac{2 n_c}{N(N-1)}$ we have $\Var(\hat p_C) = \frac{4}{N^2(N-1)^2} \langle n_C \rangle = \frac{2}{N(N-1)} p_C$. A Poisson approximation thus recovers Eq.~\ref{eqvarhatpc_poisson} identifying counting noise as the dominant source of variance in small samples.

The scaling analysis suggests that in terms of mean squared error it might be preferable to only estimate the Poisson term in the very smallest sample to reduce variance stemming from the estimation of the first two terms in Eq.~\ref{eqvarhatpc}.
Benchmarking of the Poisson estimator,
\begin{equation} \label{eqhatvarhatpc_poisson}
    \hat{\Var}_{\mathrm{Poisson}}(\hat{p_c}) = \frac{2}{N(N-1)} \hat{p_C},
\end{equation}
confirms this intuition: The Poisson estimator greatly reduces variance in small samples at the expense of moderate negative bias (Fig.~\ref{figpoisson}). In practice, we propose using the maximum value of the Poisson and unbiased estimator to increase robustness in the smallest samples.
Future work might more formally address the problem of combining the two estimators to minimize overall mean squared error using the statistical framework of shrinkage estimators \cite{Hausser2009EntropyInference}.

\begin{figure}
\begin{center}
    \includegraphics{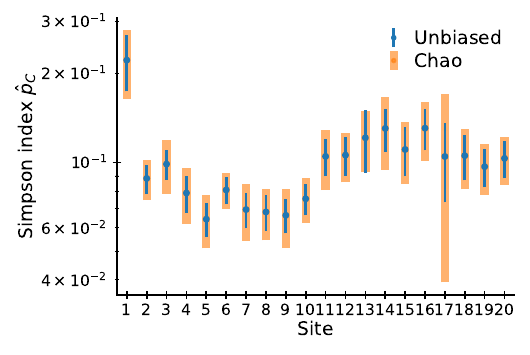}
    \caption{
        {\bf Comparison of interval estimation on empirical data.} The dataset presented in \citet{Jongman1995DataAnalysis} contains the abundances of 30 plant species at 20 different sampling sites on the Dutch island of Terschelling. Interval estimates for the biodiversity at each site were obtained from the unbiased estimator (blue) and the Chao estimator (orange) as $\hat{p_C} \pm \sqrt{\hat{\Var}_{\mathrm{method}}(\hat{p_C})}$. 
    }
    \label{figdunes}
\end{center}      
\end{figure}

\section{Comparisons on empirical data}

To demonstrate the practical importance of unbiased estimation we next compared estimators on empirical data. In a textbook dataset on how vegetation patterns depend on the management of dune meadows \cite{Jongman1995DataAnalysis} we find that the unbiased estimator produces consistently lower estimates of variance than Chao's estimator (Fig.~\ref{figdunes}). In this dataset $N$ varies from 15 to 48, which is of the same order of magnitude than the $S=30$ species which are distinguished in this dataset. The wider interval estimates of Chao's method are compatible with the previously demonstrated bias of this methodod for $N \sim S$. Overestimation of sampling variance decreases the percentage of pairs of sites with non-overlapping error bars from 44\% for the unbiased method to 29\% for Chao's method, demonstrating the potential gain in statistical power using the proposed estimator. 

\begin{figure*}[t]
\begin{center}
    \includegraphics{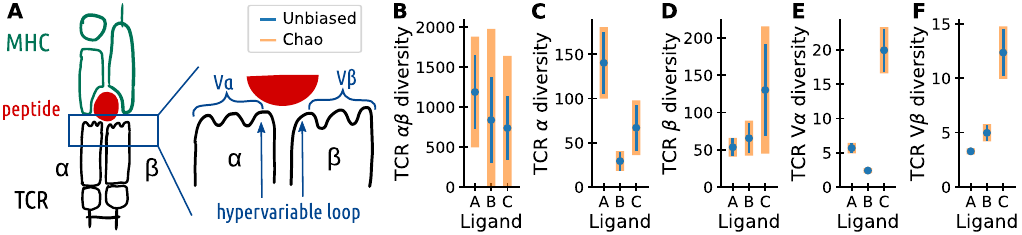}
    \caption{{\bf Quantifying the diversity of ligand-specific T cell receptors.} (A) Schematic diagram of the interaction between TCRs and their ligands, peptides bound to major histocompatibility complex (MHC).  The $\alpha\beta$TCR is a heterodimer composed of an $\alpha$ and $\beta$ chain, each containing variable loops that together determine TCR specificity to its ligand binding partners. The sequence of the variable loops are determined during genetic recombination which involves the choice of gene segments, called V genes, and additional diversification within the hypervariable complementary determining region 3. 
    (B-F) Interval estimates for the diversity of pMHC-specific receptors were obtained from the unbiased estimator (blue) and the Chao estimator (orange) for three viral epitopes. Diversity was assessed on the level of (B) the full receptor, (C) the $\alpha$ and (D) $\beta$ chain, as well as (E) the V$\alpha$ and (F) V$\beta$ gene choice.  
    Diversities are shown as effective number equivalents of Simpson's index, $D = 1/\hat{p_C}$, with error bars calculated by error propagation $\sqrt{\hat\Var(\hat p_C)}/\hat{p_C}^2$.
    Data: Dash et al. \cite{Dash2017QuantifiablePredictive}.
    Ligand A: Influenza virus peptide M1$_{\text{58}}$. Ligand B: Epstein-Barr virus peptide BMLF1$_{\text{280}}$. Ligand C: Human cytomegalovirus peptide pp65$_{\text{495}}$. The three peptides are presented by a common MHC, the Human Leukocyte Antigen (HLA) A*02:01. 
    }
    \label{figtcr}
\end{center}      
\end{figure*}

We next compared estimator performance for the problem of quantifying T cell receptor (TCR) diversity. Stochastic genetic recombination creates hypervariable TCRs, which are the molecular basis for how our adaptive immune system responds to diverse pathogens \cite{Mayer2015HowWellAdapted} (Fig.~\ref{figtcr}A). TCR diversity can be quantified by considering receptors as "species" with associated probabilities corresponding to the likelihood of clonal lineages encoding the same receptor. The number $S$ of receptors that can be created by recombination is immense with estimates as large as $S \sim 10^{39}$ for the TCR$\beta$ chain \cite{Mora2019QuantifyingLymphocyte}. Therefore this problem illustrates a practical use case for diversity estimation methods in the $N \ll S$ regime.  

To test the variance estimators we constructed a metarepertoire of 30 million T cell clonotypes by combining samples from multiple healthy donors from a cohort study \cite{Emerson2017ImmunosequencingIdentifies} and then split this repertoire into non-overlapping pools of different sizes. The unbiased estimator outperforms all other estimators in terms of bias (Fig.~\ref{figtcrrep}B) and variance (Fig.~\ref{figtcrrep}C). (Chao's estimator was excluded from this comparison due to its slow computational speed at tested sample sizes.) 
Our results show that using the unbiased estimator only >10000 sequences are needed to estimate $\Var(\hat{p_C})$ with a coefficient of variation $\ll 1$. Such sampling depths are now readily obtainable via bulk \cite{Emerson2017ImmunosequencingIdentifies} or single cell TCR sequencing \cite{Lindeboom2023HumanSARSCoV2} allowing this estimator to be applied to quantify differences in TCR diversity across individual samples.

\section{Application: How many T cell receptors bind to a given ligand?}

Having established the good empirical performance of the unbiased estimator, we sought to exploit our statistical advance to provide a quantitative answer to an important open question in immunology (Fig.~\ref{figtcr}): How degenerate is the mapping between antigen receptors and their ligands? The diversity of antigen receptors binding to a ligand determines the breadth and polyclonality of the adaptive immune response and quantification of this diversity is thus of central interest in the field \cite{Dash2017QuantifiablePredictive,Chen2017SequenceStructural,Mayer2023MeasuresEpitope}. Recent experimental advances in single cell sequencing of T cells sorted for specificity to multimerized ligands allow experimental probing of this diversity \cite{Dash2017QuantifiablePredictive,Glanville2017IdentifyingSpecificity,Minervina2022SARSCoV2Antigen}. By quantifying the diversity of ligand-specific TCRs at multiple levels, we demonstrate the statistical significance of ligand-dependent differences in the contribution of the two chains of the heterodimeric receptor to binding specificity.

The dataset we consider consists of the sequences of 415 $\alpha\beta$TCRs specific to three ligands \cite{Dash2017QuantifiablePredictive}, which are important viral epitopes (see Fig.~\ref{figtcr} caption). We label them A, B, C in the text for conciseness. Each experiment involved sorting T cells from multiple individuals, but for simplicity we will determine overall TCR diversity regardless of donor-origin, a limitation which can be relaxed as dataset sizes increase. Using our method we determined the diversity of the full receptor (Fig.~\ref{figtcr}B) and its parts (Fig.~\ref{figtcr}C-F) along with their associated sampling uncertainy. We find that the effective diversity of TCRs binding each ligand is on the order of a thousand receptors (Fig.~\ref{figtcr}B) albeit with a large associated sampling uncertainty. Interestingly, when zooming in on the diversity of the component parts of the heterodimeric receptor, we find strong statistical support for hypothesized differences \cite{Dash2017QuantifiablePredictive,Minervina2022SARSCoV2Antigen} in $\alpha$ and $\beta$ chain diversity among ligand-specific receptors (Fig.~\ref{figtcr}C-F). For instance, TCRs specific to ligand A have significantly more diverse $\alpha$ than $\beta$ chains, while the reverse is true for ligand B. The diversity of V gene segments found in specific TCRs also varies significantly between ligands and is largest among TCRs specific to ligand C.

The unbiased interval estimates (blue) are equivalent to Chao's estimates (orange) for V gene diversity (Fig.~\ref{figtcr}E-F), but are substantially tighter for the diversity estimates on the full receptor level (Fig.~\ref{figtcr}B) and receptor chain level (Fig.~\ref{figtcr}C,D), again highlighting the upward bias of alternative estimators. Importantly, the quantification of TCR diversity at different levels leads to hypotheses about the structural basis of recognition for each ligand. For example, TCRs specific to ligand C are expected to make fewer contacts on average between the V-gene encoded CDR1 and CDR2 loops and the ligand. Similarly, diversity restriction among the two chains in ligand A and B might be reflective of how many contacts each chain is making with the ligand. These hypotheses will soon become testable as more structures of TCRs in complex with their cognate ligands are solved \cite{Chen2017SequenceStructural,Song2017BroadTCR}.

\section{Conclusion and Discussion}
This work introduced a method to estimate the sample variance of Simpson's diversity index without bias for arbitrary species abundance distributions. To our knowledge the estimator we have introduced here is the only provably unbiased variance estimator for any diversity metric. This unbiased estimator does not seem to be widely known despite its superior statistical properties compared to existing methods. Additionally the unbiased estimator has a closed form analytical expression and is thus fast to calculate even for large samples, in contrast to bootstrapping approaches.

As the estimator is unbiased it has the practically important property of producing estimates that do not vary systematically with sample size. This is an important advantage for practical applications in which sample sizes vary between ecological communities. Such variation has become an increasingly important concern in ecology, as the field has moved to apply techniques developed for field studies with well-controlled sampling effort to the assessment of microbiome \cite{Haegeman2013RobustEstimation,Willis2019RarefactionAlpha} or immune repertoire \cite{Laydon2015EstimatingTcell,Kaplinsky2016RobustEstimates,Mayer2023MeasuresEpitope} diversity from high-throughput sequencing experiments. By using the unbiased estimator introduced here ecologists can avoid the loss of information inherent in the common practice of subsampling larger samples down to the smallest sample size, known as rarefaction. Our method thus fills a previously identified gap in the ecological literature \cite{Willis2019RarefactionAlpha} to overcome the need for rarefaction by bias-corrected interval estimators. 

An extension of our work could revisit methods for interval estimation for other diversity metrics such as Shannon entropy. For these metrics past work has focused on reducing bias in the point estimates themselves given the absence of an unbiased estimator \cite{Nemenman2001EntropyInference,Grassberger2003EntropyEstimates}. Our work might be generalized to address the variance estimation problem for these bias-corrected estimators for other diversity metrics. Another direction for future work is to compare the performance of the estimators on samples with overdispersion \cite{Scherer2013SimultaneousConfidence}, which goes beyond the multinomial sampling assumption that underlies all tested estimators.

We note that the negative logarithm of Simpson's index, $-\log p_C$, is the Renyi entropy (of order 2) \cite{Xu2020DiversityBiology,Roswell2021ConceptualGuide}. The Renyi entropy in turn lower-bounds Shannon entropy $-\sum_i p_i \log p_i$, a relation that has been exploited to estimate entropy rates of dynamical systems \cite{Ma1981CalculationEntropy,Nemenman2001EntropyInference} and neural spike trains \cite{Strong1998EntropyInformation}. Thus we expect that our estimator will also be of use outside of ecology in the many other areas that use the concept of entropy. Interestingly, the estimator we have introduced can determine sampling variances even when the total number of species $S$ exceeds the sample size $N$. This shows that the surprising ability to infer entropies way before the distribution is fully sampled, known in statistical physics as Ma's square-root regime of entropy estimation \cite{Ma1981CalculationEntropy,Nemenman2001EntropyInference,Hernandez2023LowprobabilityStates} and in probability theory as the birthday paradox \cite{Nunnikhoven1992BirthdayProblem,Paninski2008CoincidencebasedTest} generalizes from point to inverval estimation.

Application of the new estimator experimentally identified ligand-specific T cell receptors showed that their effective receptor diversity is on the order of $\sim1000$ and demonstrated ligand-dependent restriction of TCR chain diversity. The effective number of receptors is very small compared to the multiple trillions of $\alpha\beta$ receptors that can be produced by recombination demonstrating the stringent selection of antigen-specific TCRs in these experiments \cite{Mayer2023MeasuresEpitope}. Knowing how many TCRs on average bind a given ligand is important in experimental design for TCR screens as it can help guide the breadth and depth of sampling strategies. Quantification of variability in the effective number of TCRs binding to different ligands using the method introduced in this paper could yield insights into the mechanistic basis of immunodominance hierarchies, and help quantify how much the effective diversity of specific TCRs depends on cutoffs on TCR avidity imposed by different experimental assays, prior exposure or age \cite{Chen2017SequenceStructural,Minervina2022SARSCoV2Antigen,vandeSandt2023NewbornChildlike}. Finally, quantification of ligand-specific TCRs diversity might help predict variation in performance of machine learning models for different ligands \cite{Montemurro2022NetTCR2Lessons,Croce2023DeepLearning}.

To aid adoption of the method we have made a reference implementation of the estimator available as an open source Python package \footnote{\url{https://github.com/andim/pydiver}}. We are hopeful that our novel method for interval estimation of diversity will enable focusing of sampling efforts for monitoring biodiversity loss in our changing world.

{\bf Acknowledgements.} The author thanks Curtis G. Callan and Yuta Nagano for critical reading of the manuscript. The work was supported in parts by the Royal Free Charity and was concluded at the Aspen Center for Physics, which is supported by National Science Foundation grant PHY-2210452.

\bibstyle{apsrev4-1}
\bibliography{estimator}

\appendix

\setcounter{figure}{0}
\renewcommand{\thefigure}{S\arabic{figure}}

\section{Appendix: Derivation of the variance expression}
\label{apphatpc}

To derive the variance of Simpson's estimator we need to calculate various (cross-)moments of the multinomial distribution. Taking derivatives of the probability generating function demonstrates that the factorial moments are equal to
\begin{equation} \label{eqfactorialmoments}
    \langle \prod_i^S n_i^{(a_i)} \rangle = N^{(\sum_i^S a_i)} \prod_i^S p_i^{a_i},
\end{equation}
where $x^{(n)} = x (x-1) ... (x-n+1)$ denotes the falling factorial. To calculate the raw moments of the distribution we make use of the moment generating function,
\begin{equation}
    M(t_1, \dots, t_S) = G(e^{t_1}, \dots, e^{t_S})  = \langle e^{\sum_i^S t_i n_i} \rangle,
\end{equation}
which for the multinomial distribution is equal to
\begin{equation}
M(t_1, \dots, t_S) = (\sum_i^S p_i e^{t_i} )^N.
\end{equation}
Calculating the partial derivatives of the moment generating function at $t_1 = \dots = t_S = 0$ we obtain
\begin{align}
    \langle n_i \rangle &= N p_i \label{eqni1} \\
    \langle n_i^2 \rangle &= N^{(2)} p_i^2 + N p_i \label{eqni2} \\
    \langle n_i^3 \rangle &= N^{(3)} p_i^3 + 3 N^{(2)} p_i^2 + N p_i \label{eqni3} \\
    \langle n_i^4 \rangle &= N^{(4)} p_i^4 + 6 N^{(3)} p_i^3 + 7 N^{(2)} p_i^2 + N p_i \label{eqni4}.
\end{align}

The variance of $\hat{p_c}$ can be expressed as
\begin{equation}  \label{eqvarhatpc_def}
    \Var(\hat{p_C}) = \frac{\langle \left(\sum_i n_i (n_i-1)\right)^2 \rangle}{(N(N-1))^2} - p_C^2.
\end{equation}
The key calculation concerns the numerator of the first term, which is equal to
\begin{equation}
    \sum_i \langle (n_i^{(2)})^2 \rangle + \sum_i \sum_{j\neq i} \langle n_i^{(2)} n_j^{(2)}\rangle \\
\end{equation}
Expanding the first term, and evaluating the second average using Eq.~\ref{eqfactorialmoments} yields
\begin{equation}
    \sum_i \left( \langle n_i^4 \rangle - 2 \langle n_i^3 \rangle + \langle n_i^2 \rangle \right)
     + \sum_i \sum_{j\neq i} N^{(4)} p_i^2 p_j^2
\end{equation}
Using the expression for the moments Eqs.~\ref{eqni2}-\ref{eqni4}, and noting that $\sum_{j \neq i} p_j^2 = \sum_j p_j^2 - p_i^2 = p_C - p_i^2$, we obtain
\begin{equation} \label{eqnisq}
     4 N^{(3)} \sum_i p_i^3 + 2 N^{(2)} p_C + N^{(4)} p_C^2.
\end{equation}
Plugging this numerator into Eq.~\ref{eqvarhatpc_def} we obtain after some algebra,
\begin{equation}
    \Var(\hat{p_C}) = \frac{4 N^{(3)} \sum_i p_i^3 - 2N^{(2)}(2N-3) p_C^2 + 2 N^{(2)} p_C }{(N(N-1))^2},
\end{equation}
the expression first published by \citet{Simpson1949MeasurementDiversity}.

\begin{figure*}
\begin{center}
    \includegraphics{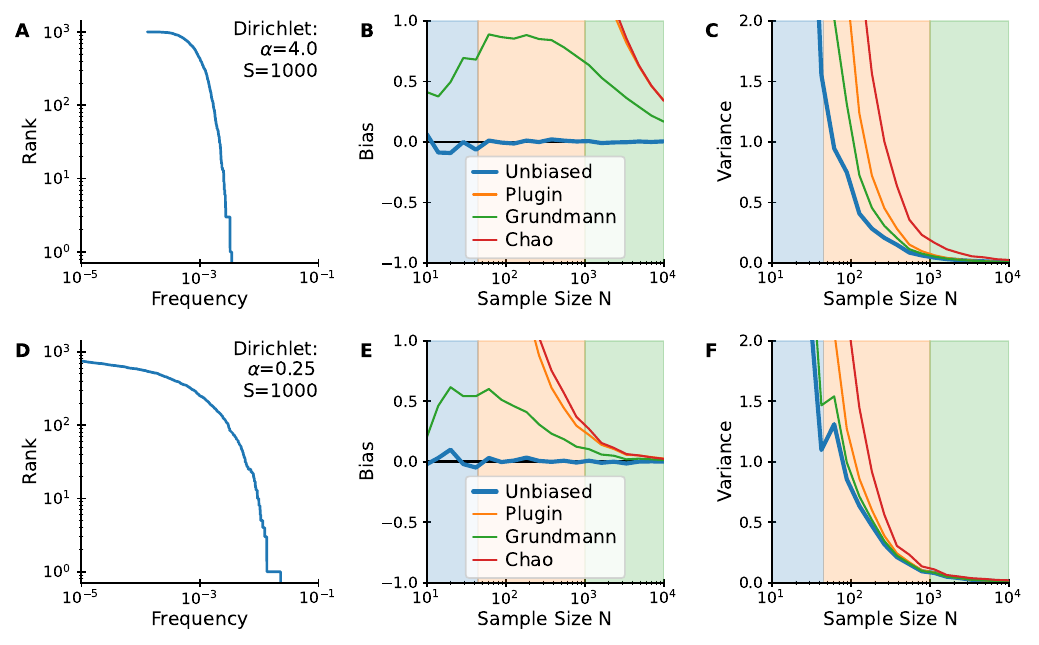}
    \caption{{\bf Supplement to Fig.~\ref{fignumerics}.} Performance of variance estimators for species abundances drawn from a Dirichlet distribution with other choices of the parameter $\alpha$. Top: $\alpha = 4.0$. Bottom: $\alpha=0.25$.
    }
    \label{fignumericsSI}
\end{center}      
\end{figure*}

\begin{figure*}
\begin{center}
    \includegraphics{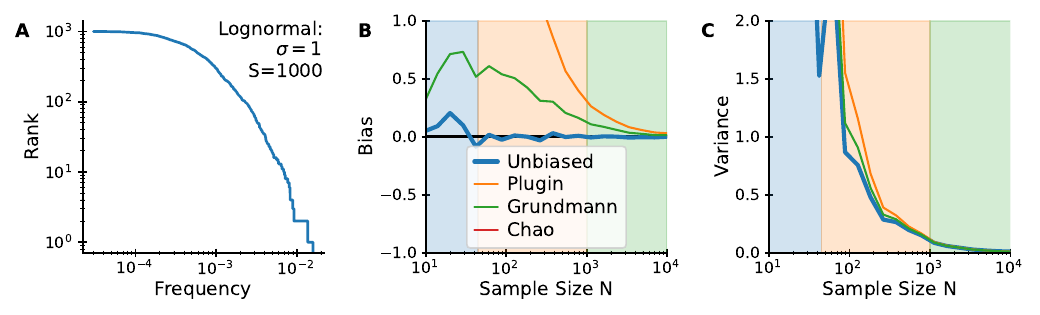}
    \caption{{\bf Supplement to Fig.~\ref{fignumerics}.} Performance of variance estimators for lognormally distributed species abundances with $\log p_i \propto N(0, \sigma)$ with $\sigma = 1$.
    }
    \label{fignumericslognormal}
\end{center}      
\end{figure*}

\begin{figure*}
\begin{center}
    \includegraphics{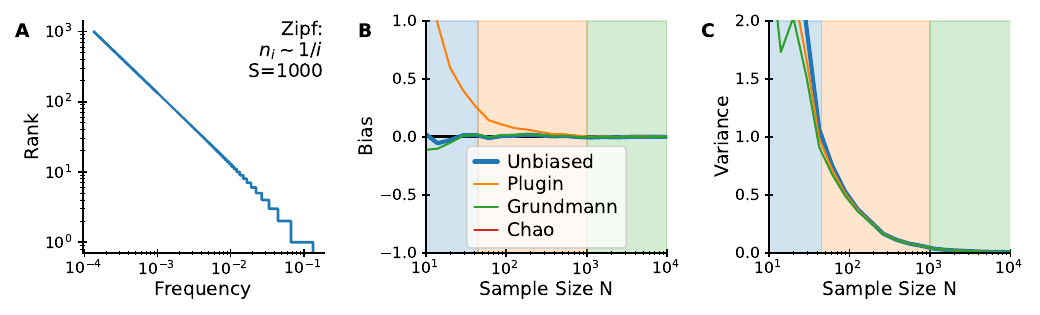}
    \caption{{\bf Supplement to Fig.~\ref{fignumerics}.} Performance of variance estimators for Zipf-distributed species abundances with $p_i \sim 1/i$.
    }
    \label{fignumericszipf}
\end{center}      
\end{figure*}

\begin{figure*}
\begin{center}
    \includegraphics{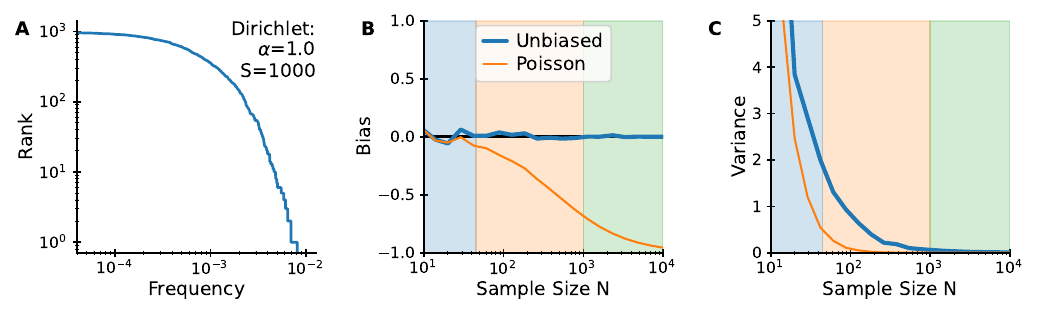}
    \caption{
        {\bf Comparison of the Poisson and unbiased estimators.} (A) Frequency-rank plot as in Fig.~\ref{fignumerics} (B) Bias and (C) variance as a function of sample size $N$ for the unbiased estimator (Eq.~\ref{eqhatvarhatpc}) and the Poisson estimator (Eq.~\ref{eqhatvarhatpc_poisson}). While the Poisson estimator is substantially negatively biased in large samples, it has only modest bias in small samples and lower variance than the unbiased estimator. 
    }
    \label{figpoisson}
\end{center}      
\end{figure*}

\begin{figure*}
\begin{center}
    \includegraphics{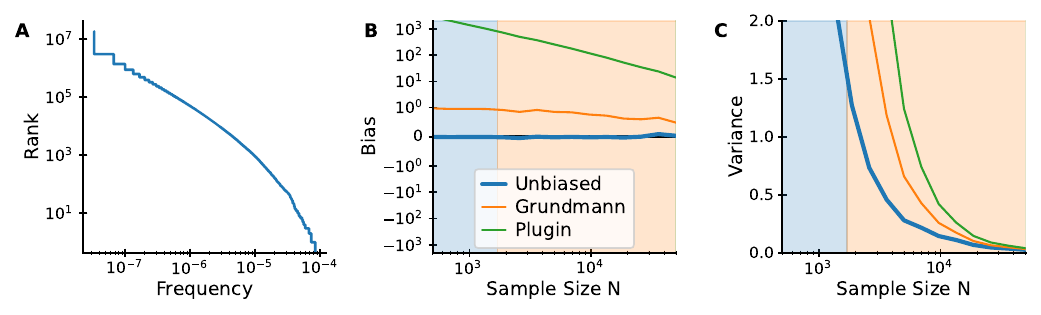}
    \caption{
        {\bf Benchmarking of variance estimators on T cell receptor repertoire data.} (A) Frequency-rank plot of TCR multiplicities among thirty million clonotypes from a metarepertoire.
         (B) Bias and (C) variance as a function of sample size $N$. Expectation values were calculated at each sample size by splitting the total sequence pool into non-overlapping subsets. Bias and variance are expressed as fractions, i.e. divided by the true value or its square, respectively. The shaded areas differentiate sampling regimes: blue $N < \sqrt{2D}$, orange $\sqrt{2D} < N $ with effective diversity $D = 1/\hat{p_C} \sim 1.5 \cdot 10^6$ (estimated by applying Eq.~\ref{eqhatpc} to the complete dataset). Note that in (B) a logarithmic scale is used for absolute values larger than 1 to account for the large bias of the plugin method. 
         Data: A metarepertoire of 30 million complementary determining region 3 (CDR3) sequences of the TCR $\beta$-chain was constructed by random selection from the combined productive clonotypes of 200 healthy donors from the \citet{Emerson2017ImmunosequencingIdentifies} cohort study.
    }
    \label{figtcrrep}
\end{center}      
\end{figure*}

\end{document}